\documentclass[12pt]{article}
\usepackage{amsxtra,amssymb,amsthm,amsmath,latexsym}

\theoremstyle{plain}

\newcommand{\refS}[1]{Section~\ref{S:#1}}

\def\R{{\mathbb R}}
\def\C{{\mathbb C}}
\def\calU{{\mathcal U}}
\def\tildet{{\widetilde\tau}}
\def\tildeD{{\widetilde D}}
\def\diam{{\hbox{\,diam\,}}}
\def\dist{{\hbox{\,dist\,}}}

\def\bee{\begin{equation*}}
\def\eee{\end{equation*}}
\def\be{\begin{equation}}
\def\ee{\end{equation}}

\begin{document}
\title{ Wave scattering by small particles in a medium.}

\author{A.G. Ramm\\
 Mathematics Department, Kansas State University, \\
 Manhattan, KS 66506-2602, USA\\
ramm@math.ksu.edu}

\date{}
\maketitle\thispagestyle{empty}
\begin{abstract}
\footnote{key words: wave scattering, small particles, many-body 
scattering problem}

Wave scattering is considered in a medium in which many small particles are
embedded. Equations for the effective field in the medium are derived
when the number of particles tends to
infinity.

\end{abstract}

\section{Introduction}\label{S:1}

Assume that waves in the medium are described by the equation
\be\label{e1}
 Lu:=[(a_{ip}(x)u_{,p})_i+k^2 n(x)]u=0 \hbox{\quad in\quad}\R^3, \ee
where over the repeated indices summation is understood,
$u_{,p}:=\frac {\partial u}{\partial x_p}$,
and the Green function, satisfying the radiation condition, solves the 
equation:
\be\label{e2} LG=-\delta(x-y) \hbox{\quad in\quad}\R^3. \ee
If there are $M$ particles $D_m$, placed in the medium, situated in a 
bounded domain $D$, 
outside of which 
\be\label{e3}
 a_{ip} (x)=\delta_{ip},\quad n(x)=1,\quad x\in D':=\R^3\setminus D, \ee
where $a_{ip}(x)$, $n(x)$, are $C^2\hbox{-smooth}$ functions,
$\delta_{ip}$ is the Kronecker symbol,
and the ellipticity condition holds: 
\bee
 c_1\sum^3_{p=1} |t_p|^2 \leq \sum^3_{c,p=1} a_{ip} t_p\overline{t_i}
 \leq c_2\sum^3_{p=1} |t_p|^2, \quad c_1>0,\eee
where $t\in \C^3$ is an arbitrary vector,
then the scattering problem consists of solving the equation
\be\label{e4}
  L\calU=0 \hbox{\quad in\quad} \R^3\backslash\bigcup_{m=1}^MD_m, \ee
\be\label{e5}
 \calU\big|_S=0 \hbox{\quad on\quad}S_m,\quad 1\leq m\leq M,\ee
\be\label{e6}
 \calU=\calU_0+\sum^M_{m=1} \int_{S_m} G(x,s)\sigma_m(s)\,ds.\ee Here
$L\calU_0=0$, $u_0$ is the scattering solution in the absence of
particles, i.e., if $M=0$.

By Ramm's lemma \cite{R470}, p. 257, one can define the scattering 
solution $ \calU_0$ in the absence of small particles by the relation:
\be\label{e7}
 G(x,y)=g(y) \calU_0(x,\beta)[1+o(1)], \quad |y|\to\infty,
 \quad \frac{y}{|y|}=-\beta,\ee
where $g(y):=\frac{e^{ik|y|}}{4\pi|y|}$.
We assume that $ka<<1$, where 
$a=\frac12 \max_{1\leq m\leq M}\diam D_m$.

The aim of this paper is to develop a general approach to wave scattering in 
a medium in which many small particles are embedded. Smallness of the 
particles is understood in the sense $ka<<1$. 
The functions $n(x)$ and $a_{ip}(x)$ are assumed practically constant on 
the 
scale of the wavelength
$$k(|\nabla n|+|\nabla a_{ip}|)<<1.$$
We generalize the approach developed in \cite{R476}, \cite{R518}, 
\cite{R506}, \cite{R450}. Earlier works are \cite{F}, \cite{R}, 
\cite{V}, to mention a few.

Our basic result is a formula for the wave field in the medium in which 
small particles are embedded. This field solves equation 
\eqref{e4}--\eqref{e6} and satisfies the radiation condition at infinity:
\be\label{e8}
 \frac{\partial(\calU-\calU_0)}{\partial r} - 
ik(\calU-\calU_0)=o\left(\frac{1}{r}\right), 
\quad r:=|x|\to\infty. \ee
We assume that
\be\label{e9}  d\gg a, \quad ka\ll 1, \ee
where $d=\min_{m\not= j}\dist(D_m,D_j)$, and the $\dist$ denotes the distance 
between two sets. Near any point $x\in D$, such that 
$\min_{1\leq m\leq M}\dist(x,D_m)>>1$, 
one calculates the wavefield $\calU$ by the formula
\be\label{e10}
 \calU(x)=\calU_0(x)+\sum^M_{m=1} G(x,x_m) 
 \int_{S_m} [1+ik\nu\cdot(s-x_m)]\sigma_m(s)ds, \ee
where $\frac{\nu}{|\nu|}\approx\frac{x-x_m}{|x-x_m|}$, and 
$\nu$ depends on $x_m,x,$ and on the functions $n(x)$, $a_{ip}(x)$,
The term $ik\nu\cdot(s-x_m)$ comes from the formulas
\bee\begin{aligned}
 \int_{S_m} &
 G(x,s)\sigma_m(s)ds =G(x,x_m)
 \left\{ \int_{S_m} \sigma_m(s)ds+ \int_{S_m} 
 \frac{[G(x,s)-G(x,x_m)]}{G(x,x_m)}ds \right\},\\
 & \frac{G(x,s)-G(x,x_m)}{G(x,x_m)}
  =\frac{\int^1_0\nabla_yG(x,x_m+\tau(s-x_m))\cdot(s-x_m)d\tau}{G(x,x_m)}
  :=ik\nu\cdot(s-x_m),\end{aligned} \eee
where $\nu=\nu(x,x_m)$, $\frac{\nu}{|\nu|}=\frac{x_m-x}{|x_m-x|}$.
One may consider $\nu$ as a known vector because $G(x,y)$ is known.

In a generic case, when $Q_m:=\int_{S_m}\sigma_m(s)ds\not=0$, the 
assumption
$ka<<1$ allows one to neglect the term $ik\nu\cdot s=O(ka)$ in \eqref{e10}
and to write \eqref{e10} as
\be\label{e11}
 \calU(x)=\calU_0(x)+\sum^M_{m=1} G(x,x_m)Q_m,
 \quad Q_m=\int_{S_m} \sigma_m ds, \ee
If $Q_m=0$, then the term $ik\nu\cdot s$ cannot be neglected. 
We discuss this case in \refS{3}. A physical example of such a case is 
the scattering by acoustically hard particles when the boundary condition
is the Neumann one: $\frac{\partial \calU}{\partial N}\big|_{S_m}=0$, 
$1\leq m\leq M$.


\section{General methodology}\label{S:2}

Let us first assume that \eqref{e11} is applicable and calculate $Q_m$. In a 
neighborhood of $S_j$ one has the exact boundary condition \eqref{e5}, which 
can be written as:
\be\label{e12}
 \int_{S_j} G(s,t)\sigma_j(t)dt 
 =-\left(\calU_0(s)+\sum_{m\not= j} G(s,x_m)Q_m\right):=u_e(s),\ee
where $s\in S_j$ and $u_e$ is the effective field acting on $D_j$. 
The basic assumption is: 

{\it We assume that $u_e(s)$ is practically constant on the 
distances of order $a$.} 

As $|s-t|\to 0$, one has
\be\label{e13}
 G(s,t)=\frac{1}{4\pi|s-t|}\ [1+O(ka)], \quad |s-t|\to 0,\ee
where $s,t\in S_j$ and we have assumed for simplicity that 
$a_{ij}=\delta_{ij}$.
In the general case one replaces the function 
$g_0(s,t):=\frac{1}{4\pi|s-t|}$ on the surface $S_j$ by the fundamental 
solution of the operator 
$\sum^3_{i,p=1}a_{ip}(x_j)\frac{\partial^2}{\partial x_i\partial x_p}$,
which can be written explicitly and analytically:
\bee
 G(x,y)=\frac{1}{4\pi\sqrt{det(a_{ip})} }
 \ \frac{1}{ [a^{(-1)}_{ip}(x_i-y_i)(x_p-y_p)]^{1/2} }, \eee
where the matrix $a^{(-1)}_{ip}$ is inverse of the matrix $a_{ip}(x)$,
$x_i$ is the $i\hbox{-th}$ Cartesian component of the vector $x$ (not to 
be confused with the vector $x_j=x$).

If $a_{ip}=\delta_{ip}$, then $G$ solves the equation
\be\label{e14}
 G(x,y)=g_0(x,y)+k^2 \int_D g(x,t) n(t) G(t,y)dt,
 \quad  g_0(x,y)=\frac{1}{4\pi|x-y|}.\ee
The integral in \eqref{e14} is $O(1)$ as $|x-y|\to 0$, so \eqref{e13} 
follows from \eqref{e14}. We have:
\be\label{e15}
 G(x,y)=g_0(x,y)[1+O(|x-y|)], \qquad |x-y|\to 0.\ee
In this paper $k>0$ is assumed fixed.  
The error term in \eqref{e13} is $O(ka)$, and 
the small parameter $ka$ is dimensionless. Replacing $G(s,t)$ in 
\eqref{e12} 
by its expression  \eqref{e13} and neglecting the small term $O(ka)$, one 
gets 
the following
integral equation for $\sigma_j$:
\be\label{e16}
 \int_{S_j} g_0(s,t)\sigma_j(t)dt=-u_e(x_j), \ee
where we may replace $u_e(s)$, $s\in S_j$, by the quantity $u_e(x_j)$ 
because 
$|x_j-s|\leq a$, and $u_e$ was assumed practically constant on the 
distance of order $O(a)$.
Equation \eqref{e16} is an equation for the electrostatic charge density 
$\sigma_j(t)$ on 
the surface of a perfect conductor $D_j$, charged to a constant potential 
$-u_e(x_j)$. The total charge on the surface of this 
conductor is
$$Q_j=\int_{S_j}\sigma_j dt.$$ 
One knows 
from electrostatics that the  total charge on the surface of a
perfect conductor $D_j$ equals to the product of 
the electrical capacitance $C_j$ of this conductor and the potential 
$-u_e(x_j)$, to which this conductor is charged:
\be\label{e17} Q_j=-C_j u_e(x_j). \ee

\textit{
In \cite[p.26, formula 5.12]{R476}, analytical formulas are derived for 
calculating the electrical capacitances of conductors of arbitrary 
shapes with any desired accuracy.}

Thus, one may consider the capacitance $C_j$ to be known. Equation 
\eqref{e11} 
can be written as 
\be\label{e18}
 \calU(x)=\calU_0(x)-\sum^M_{m=1} G(x,x_m) C_m\calU(x_m),
 \quad \dist(x,D_m)\gg a.\ee
Equation \eqref{e17} can be considered as a linear algebraic system for 
finding $Q_m$. Namely, \eqref{e12} and \eqref{e17} imply:
\be\label{e19}
 Q_j=-C_j\left(\calU_0(x_j)+\sum_{m\not= j} G(x_j,x_m)Q_m\right).\ee
Equation \eqref{e19} is a linear algebraic system for the $M$ unknowns $Q_j$.
The matrix elements of this sytem are known because the Green's function
 $ G(x, y)$ is known.
If the condition
\be\label{e20}
 C_j \sum_{m\not= j} |G(x_j,x_m)|<1,\ee
holds, then system \eqref{e19} can be solved by iterations and the 
iterative 
process
\be\label{e21}
 Q^{(n+1)}_j =-C_j\calU_0(x_j) -\sum_{m\not= j} C_j G(x_j,x_m)Q^{(n)}_m,
 \quad Q^{(0)}_m:=-C_j \calU_0(x_j), \ee
converges to $Q_j$ at the rate of a geometric series as $n\to \infty$.
If $M\to\infty$ and the limit
\be\label{e22}
 \int_\tildeD C(y)dy=\lim_{M\to\infty} \sum_{D_j\subset \tildeD}C_j \ee
exists for any subdomain $\tildeD\subset D$, where the function 
$C(y)$ is integrable, then this function is the limiting density of the 
electrical capacitance of the small 
particles. If the limit \eqref{e22} exists, then equation \eqref{e18} in this 
limit takes the form:
\be\label{e23}
 \calU(x)=\calU_0(x)-\int_D G(x,y)C(y)\calU(y)dy.\ee
Here $\calU(x)$ is the limit of $u_e(x)$ as $M\to\infty$ and
it is assumed that \eqref{e22} 
holds.

Applying the operator $\nabla+k^2n(x)$ to \eqref{e23} and using
equation \eqref{e2}  yields the following equation:
\be\label{e24}
 [\nabla^2+k^2n(x)]\calU=C(x)\calU(x). \ee
This is a Schr\"odinger-type equation for the effective (self-consistent) wave 
field $\calU$:
\be\label{e25}
 [\nabla^2+k^2-q(x)]\calU=0, \ee
where
\be\label{e26}
 q(x):=C(x)+k^2[1-n(x)].\ee
Thus, in the limit $M\to \infty$ under the assumption \eqref{e22},
we have derived a linear Schr\"odinger-type equation \eqref{e25} for the 
effective wave field $\calU$. If the small particles are identical, the 
electrical capacitance of a single conductor with the shape of a particle is 
$C_0$, and the density of the number of small particles is $N(x)$,
then $C(x)=N(x)C_0$, so $N(x)$ can be calculated if $C(x)$ and $C_0$ are 
known. By the density $N(x)$ of the number of small particles
we mean the function  $N(x)$, defined by the formula, similar to
\eqref{e22}:
\bee
 \int_\tildeD N(x)dx=\lim_{M\to\infty} \sum_{D_j\subset \tildeD} 1.
\eee

Let us summarize our method for calculating the effective field in the 
medium in which $M$ small particles are embedded and the Dirichlet condition 
\eqref{e5} holds on their surfaces.

This field is calculated by formula \eqref{e11} at any point $x$ such that
$\min + m \dist(x,D_m) 
>> a$, and the unknown numbers $Q_m$ are 
calculated
by solving linear algebraic system \eqref{e19}. If condition \eqref{e20}
holds, then system \eqref{e19} is uniquely solvable by iterations.
Condition \eqref{e20} always holds if $M$ is fixed and $a$ is sufficiently
small, because $C_m=O(a)$.

If $M\to\infty$ and condition \eqref{e22} holds, then the field $\calU$ 
can be found from the integral equation \eqref{e23}. Solving this equation 
is equivalent to solving equation \eqref{e25} where $\calU-\calU_0$ 
satisfies \eqref{e8}.
The function $n(x)$ is known, therefore $G(x,y)$ is also known.

\section{The case when $Q_m=0$.}\label{S:3}

Consider the same problem as in \refS{2} but with the Neumann boundary 
condition
\be\label{e27}
 \frac{\partial\calU}{\partial N}\bigg|_{S_m}=0, \quad 1\leq m\leq M,\ee
in place of the Dirichlet one. Here $N$ is the unit exterior (i.~e.~, 
pointing outside of $D_m$) normal to $S_m$, $1\leq m\leq M$.
We look for the solution of the same form \eqref{e6}, and the methodology is 
the same, but now $Q_m=0$, $1\leq m\leq M$, as we show below. Thus, the term 
$ik\nu\cdot s$ in \eqref{e10} becomes important. Arguing as in \refS{2}, we 
obtain in place of \eqref{e12} the following equation
\be\label{e28}
 \sigma_j(s)=A_j\sigma_j+2\frac{\partial u_e(s)}{\partial N}, 
 \quad s\in S_j,\ee
where
\be\label{e29}
 A_j\sigma_j:=2 \int_{S_j} 
 \ \frac{\partial G(s,t)}{\partial N_s}\,\sigma_j(t)\,dt,\ee
and the known formula from the potential theory:
\be\label{e30}
 \frac{\partial}{\partial N^-_s} \int_{S_j} G(x,t)\sigma_j(t)dt
 =\frac{A_j\sigma_j-\sigma_j}{2}, \ee
was used.
Here $\frac{\partial}{\partial N^-_s}$ is the normal derivative on $S_j$
from the outside of $D_j$. 

If $|s-t|\to 0$, then 
$$G(s,t)\approx g_0(s,t)=\frac{1}{4\pi|s-t|},$$ 
where
we again assume for simplicity that $a_{ip}=\delta_{ip}$. We have
\be\label{e31}
 G(x_j,s)=G(x_j,x_m)
 \left[ 1+\frac{G(x_j,s)-G(x_j,x_m)}{G(x_j,x_m)}\right]_,
 \quad s\in S_m. \ee
Since $|s-x_m|\leq a$ and $|x_m-x_j|\gg a$, we have
\bee
 \frac{G(x_j,s)-G(x_j,x_m)}{G(x_j,x_m)}
 =\frac{\int^1_0
	\frac{ d\,G(x_j,\, x_m+\tau(s-x_m))}{d\tau}d\tau}{G(x_j,x_m)}
 = \frac{\int^1_0 d\tau\nabla_yG(x_j,y)
	\big|_{y=x_m+\tau(s-x_m)}\cdot(s-x_m)}
        {G(x_j,x_m)}. \eee
Thus,
\be\label{e32}
 \frac{G(x_j,s)-G(x_j,x_m)}{G(x_j,x_m)}\approx ik \nu\cdot(s-x_m),\ee
where, by the mean value theorem, one has
\be\label{e33}
 ik\nu=\frac{\nabla_yG(x_j,x_m+\tildet(s-x_m))}{G(x_j,x_m)},
 \quad 0<\tildet<1. \ee
If $|x_j-x_m|>>a$, then
\be\label{e34}
 \frac{\nu}{|\nu|}\approx \frac{x_m-x_j}{|x_m-x_j|},\ee
and $|\nu|$ depends on the functions $n$ and $a_{ip}$. Since $n(x)$ is 
known, then $G(x,y)$ is known, so $\nu(x_m)$ can be considered as known.

If $a_{ip}=\delta_{ip}$, then the integral equation for $G$ can be written 
as
\be\label{e35}
 G(x,y)=g(x,y) -\int_D g(x,\xi)q(\xi)G(\xi,y)d\xi
 :=g-TG, \ee
where $T$ is the integral operator,  defined in \eqref{e35}, 
$$g(x,y):=\frac{e^{ik|x-y|}}{4\pi|x-y|}, \quad q(y):=k^2[1-n(y)],\quad 
q=0 \hbox{\quad in \quad} D':=\R^3\backslash D.$$ 
The operator $T$ is compact in
$L^2(D)$.

The function 
$$\frac{\nabla_y g}{g}\approx ik\frac{y-x}{|y-x|} \hbox{ \quad if 
\quad} k|y-x|\gg 1.$$
More precisely, we have used the exact formula:
$$\nabla_y g=g(x,y) \big(ik-\frac{1}{|y-x|}\big) \frac{y-x}{|y-x|},$$ 
and have neglected the term $\frac{1}{|x-y|}$ compared with $|ik|=k$. 

This is justified if $k|x-y|\gg 1$. 

Equation \eqref{e35} is uniquely solvable 
in $L^2(D)$ because it is of Fredholm type and its homogeneous version has 
only the trivial solution.
Indeed, if $h$ solves the homogeneous equation \eqref{e35},
then $$(\nabla^2+k^2-q(x))h=0 \quad \hbox{\,\, in\,\,} \R^3,$$
where the potential $q$ is 
compactly supported, $k^2>0$, 
and $h$ satisfies the radiation condition. It is known that this implies 
$h=0$. In Kato's paper
\cite{K} a similar but much stronger result is obtained: it is not
assumed that $q$ is compactly supported, the potential
may staisfy the assumption $|q(x)|=o(|x|^{-1})$ as $|x|\to \infty$.

The unique solution to \eqref{e35} has the form 
$$G=(I+T_1)g,$$ 
where  $I$ is the identity operator and
$T_1:=(I+T)^{-1}-I$ is a linear compact operator in $L^2(D)$.

Therefore 
$$\frac{\nabla_y G}{G} = \frac{(I+T_1)\nabla_y g}{G(x,y)}.$$
If $|x-y|\to\infty$ and $k>0$ is fixed, then 
\be\label{e36}
 \frac{G(x,y)}{g(x,y)}=1+O\left(\frac{1}{|x-y|}\right)
 \qquad \dist(x,D)\gg 1, \quad \dist(y,D)\gg 1. \ee
This follows from equation \eqref{e35} if one takes into account the following
estimates:
$$|g(x,\xi)|=O\left(\frac{1}{|x-\xi|}\right)\quad \hbox{\,\, as \,\,}
|x-\xi|\to\infty,$$  
$$|G(\xi,y)|=O\left(\frac{1}{|\xi-y|}\right)\quad \hbox{\,\, as \,\,}
|\xi-y|\to\infty,$$ 
and
\be\label{e37}
 J:=\int_D \frac{dz}{|x-z||z-y|} = O\left(\frac{1}{|x-y|}\right), 
 \quad |x-y|\to\infty. \ee
The last estimate holds if $D\subset\R^3$ is a bounded domain.
To prove estimate \eqref{e37}, take the origin at the point $y$, note that 
\eqref{e37} holds if $\dist(y,D)\leq\diam D$, and $J$ decays when 
$\dist(y,D)$ grows remaining less than $\frac{|x-y|}{2}$, if the gravity 
center of $D$ moves along the line joining $x$ and $y$.

If $x$ or $y$ remain in $D$ and $|x-y|\to\infty$, then the relation 
$\frac{G(x,y)}{g(x,y)}\approx 1$ does not hold, in general.

Since $G(x,y)$ is known one may assume 
that $|\nu|=|\nu(x)|=|\nu(x,y)|$ is known.

Because of \eqref{e13}, the operator $A_j$ in equation \eqref{e28} can be 
approximated by the operator 
$$A\sigma= \int_{S_j} \frac{\partial g_0(s,t)}{\partial N_s}
\sigma(t)dt.$$
It is known (see \cite[p.96, formula (7.21)]{R476}) that
\be\label{e38}
 \int_{S_j} A\sigma ds=-\int_{S_j}\sigma ds.
\ee
Therefore equation \eqref{e28} implies
\be\label{e39}
 \int_{S_j}\sigma_j ds=\int_{S_j} \frac{\partial u_e}{\partial N} 
 ds=\int_{D_j}\Delta u_e dx\approx V_j\Delta u_e(x_j),
 \quad V_j:=|D_j|,\ee
where $|D_j|$ is the volume of $D_j$. We had assumed that $u_e(x)$ is 
$C^2\hbox{-smooth}$.
The term $V_j=O(a^3)$, and  $\Delta u_e=O(k^2)$, because 
$\Delta \calU_0=-k^2n(x)\calU_0=O(k^2)$.
Thus, $$V_j\Delta u_e=O(k^2a^3).$$ Let us show that the second term in 
\eqref{e10}, namely,  $ik\int_{S_j}\nu\cdot(s-x_j)\sigma ds,$ is of the same 
order of magnitude $O(k^2a^3)$.

We have assumed that $\nu$ is practically constant on the scale of order 
$O(a)$. Thus,
\be\label{e40}
 ik\sum^3_{q=1} \nu_q \int_{S_j} (s-x_j)_q \sigma_j(s) ds
 = -ik\sum^3_{q=1,p=1}\nu_q V_j\beta^{(j)}_{qp}
 \frac{\partial u_e}{\partial x_p}.\ee 
Here we have used the following result (see \cite[p.98]{R476}):

If
\bee
  h_p=Ah_p-2N_p,
\eee
then 
\bee
\int_{S_j}(s-x_j)_q h_p 
  ds=V_j\beta^{(j)}_{pq},
\eee
where $(s-x_j)_q$ is the $q-$th Cartesian component of the vector
$s-x_j$, $V_j$ is 
the volume of the domain $D_j$, and $\beta^{(j)}_{pq}$ is the magnetic 
polarizability tensor, defined in \cite[p.62, formula (5.62)]{R476}. In 
the cited formula
one takes the magnetic constant $\mu_0=1,$ and the origin is at the 
point $x_j$, the gravity 
center of $D_j$.
The right-hand side of \eqref{e40} is of order $O(a^3)=O(k^2a^3)$ because $k>0$ 
is fixed. Thus, if the Neumann boundary condition holds on $S_m$, $1\leq 
m\leq M$, then the terms $\int_{S_m}\sigma_mds$ and 
$ik\int_{S_m}\nu\cdot(s-x_m)ds$ in \eqref{e10} are of the same order of 
smallness as $ka\to 0$,  and $k>0$ is assumed fixed.

Let us compare this conclusion with the case when the Dirichlet boundary 
condition holds on $S_m$, $1\leq m\leq M$. In this case, as follows from 
\eqref{e17}, $Q_m=O(a)$, because $C_m=O(a)$. If $a\to 0$, then 
$O(a)\gg O(a^3)$. This is the reason for the different physical 
conclusions in 
two cases. In the case of the Dirichlet boundary condition the scattering 
amplitude is of order $O(a)$ and the scattering by a single small particle 
is isotropic, while in the case of the Neumann boundary condition the 
scattering amplitude is of order $O(k^2a^3)$ and the scattering by a 
single small particle is anisotropic.

Let $M\to\infty$ in the case of the Neumann boundary condition, and assume 
$a_{ip}=\delta_{ip}$. Equations \eqref{e10}, \eqref{e39} and \eqref{e40} 
yield
\be\label{e41}
 \calU(x)=\calU_0(x)+ \int_D G(x,y)
 \left[\Delta\calU(y)-ik\sum^3_{p,q=1} \nu_q(x,y)
 \frac{\partial\calU}{\partial y_p} \beta_{pq}(y)\right] v(y)dy. \ee
Here we have assumed that for any subdomain $\tildeD\subset D$ one has:
\be\label{e42}
 \int_\tildeD v(y)dy =\lim_{M\to\infty} \sum_{D_j\subset\tildeD}V_j,\ee
\bee
 \int_{\tildeD} \beta_{pq}(y)dy=\lim_{M\to\infty}\sum_{D_j\subset\tildeD}
 \beta_{pq}^{(j)},\eee
and then we have passed to the limit $M\to\infty$ in equation \eqref{e10}.

In \cite[p.55, formula (5.15)]{R476}, analytical formulas are given for 
calculating tensor $\beta^{(j)}_{pq}$ with any desired accuracy for a body 
$D_j$ of arbitrary shape.

\section{Conclusion.}\label{S:4}
The methodology we have developed for solving many-body wave scattering 
problem for small particles, embedded in a known medium, has the following 
new features:

1) Scattering by small particles is considered in a medium.

2) If the number $M$ of the particles is  not very large, the scattering 
problem is reduced to solving linear algebraic systems with matrices, 
whose entries have physical meaning. Analytical formulas for calculating 
these entries are obtained in \cite{R476}. The reduction to linear 
algebraic systems bypasses any usage of integral equations, which are 
usually serving as a basic tool in the scattering theory.

3) If $M\to\infty$, then some integral equations (equations \eqref{e23} 
and \eqref{e41}) are derived for the effective field in the medium in 
which 
small particles are embedded.

4) In the case of the Dirichlet boundary condition the relative volume of 
the embedded particles tends to zero as $a\to 0$, $M\to\infty$.

Indeed, the number of small particles per unit volume of the medium is 
$O(\frac{1}{d})$, where $d$ is defined in \eqref{e9}, the volume of a 
single particle is $O(a^3)$, so the relative volume of the small particles 
is $O(\frac{a^3}{d})\to 0$ as $M\to\infty$, since $a\to 0$ if 
$M\to\infty$.

In the case of the Neumann boundary condition the relative volume of the 
small particles tends to a finite non-zero limit $v(y)$ (cf. \eqref{e42}).

5) The methodology, developed in this paper, can be used in the problems 
of electromagnetic wave scattering by small particles embedded in a known 
medium. It can also be used in some nanotechnological problems, consisting 
of creating "smart" materials with the desired properties, for example,
wave-focusing properties  
(see \cite{R515}, \cite{R516}).


\end{document}